\documentclass[prb,aps,twocolumn,showpacs,superscriptaddress,floatfix,amsmath,amssymb]{revtex4}
\usepackage{graphicx}
\usepackage[dvips]{color}

\begin{document}
\title{Coexistence of superconductivity and antiferromagnetism in self-doped bilayer $t$-$t'$-$J$ model}

\author{J. Y. Gan}
\affiliation{Institute for Materials Research, Tohoku University,
Sendai 980-8577}

\author{M. Mori}
\affiliation{Institute for Materials Research, Tohoku University,
Sendai 980-8577}

\author{T. K. Lee}
\affiliation{Institute of Physics, Academia Sinica, Nankang,
Taipei, Taiwan 11529}

\author{S. Maekawa}
\affiliation{Institute for Materials Research, Tohoku University,
Sendai 980-8577}
\begin{abstract}
A self-doped bilayer $t$-$t'$-$J$ model of an electron- and a hole-doped planes is studied by the slave-boson mean-field theory.
A hopping integral between the differently doped planes, which are generated by a site potential, are renormalized by the electron-electron correlation.
We find coexistent phases of antiferromagnetic (AFM) and superconducting orders, although the magnitudes of order parameters become more dissimilar in the bilayer away from half-filling.
Fermi surfaces (FS's) with the AFM order show two pockets around the nodal and the anti-nodal regions.
These results look like a composite of electron- and hole-doped FS's.
In the nodal direction, the FS splitting is absent even in the bilayer system, since one band is flat due to the AFM order.
\end{abstract}


\date{\today}

\maketitle
\section{Introduction} \label{sec1}

High-$T_c$ superconductors (HTSC) have one or more CuO$_2$ planes in a conducting block, which is separated by charge-reservoir blocks.
In HTSC with more than three CuO$_2$ planes in a unit cell, there exist two inequivalent types of CuO$_2$ planes; pyramidally-coordinated-outer planes (OP) and square-coordinated-inner planes (IP).
The nuclear magnetic resonance (NMR) studies found that the hole density in OP is lager than that in IP~\cite{Toku00,Kote01,Kote04,Muku06,Muku06JPSJ,Iyo07}.
An example of these is the five-layered HgBa$_2$Ca$_4$Cu$_5$O$_y$, in which the optimally doped OPs are superconducting (SC) with $T_c=108K$, while the three IPs have an antiferromagnetic (AFM) moment~\cite{Kote04,Muku06,Muku06JPSJ}.
Although the SC planes are separated by the AFM ones, the Josephson coupling through the AFM planes stabilizes the superconductivity as a bulk~\cite{Mori05}.

Another kind of multilayered HTSC is the four-layered Ba$_2$Ca$_3$Cu$_4$O$_8$(O$_x$F$_{1-x}$)$_2$ (F0234)~\cite{Iyo03,Iyo04}.
Especially for $x=0$, a nominal Cu valence is $+2$ on the canonical chemical formula.
Thus, this material is expected to be a Mott insulator, although the superconductivity with $T_{\rm c}$=60K takes place~\cite{Iyo03,Iyo04}.
This compound, F0234, has four CuO$_2$ planes, among which two OP's have apical F atoms, while the two IP's do not.
Angle-resolved-photoemission-spectroscopy (ARPES) experiments observed two Fermi surfaces (FS's), whose volumes in the first Brillouin zone correspond to electron- and hole-doped FS's~\cite{Chen06,Xie07}.
This would be the first self-doped high $T_c$ superconductor with an electron- and a hole-doped CuO$_2$ planes in the same crystal.
It is also found that the superconducting (SC) gap on the electron-doped FS is twice as large as that on the hole-doped one~\cite{Chen06}.

On the other hand, it is known that doped holes make a FS around the nodal region~\cite{Ino02,Yosh03,Ronn03,Shen05}, while doped electrons create pockets around the anti-nodal regions~\cite
{Armi02}.
Theoretical studies by the variational Monte Carlo method~\cite{TKLee97,TKLee03}and the exact diagonalization method~\cite{Tohy00,Tohy04} elucidate that the {\it asymmetry} between hole- and electron-doped cuprates results from second neighbor hopping ($t'$) and third neighbor one ($t''$) in the CuO$_2$ plane.
Here, the question arises; what is the ground state of the self-doped bilayer cuprates, where one plane is electron-doped and the other is hole-doped, and how are the FS's and their {\it asymmetry}?

To answer these questions, the self-doped $t$-$t^\prime$-$J$ model is examined by the slave-boson mean-field theory.
The two different types of planes are connected by an interlayer hopping renormalized by electron-electron correlation.
A site potential making the charge imbalance between two planes is included.
Note that the hopping of a single spin between a holon- and a doublon-sites picks up extra {\it minus} sign as compared to that between a holon- and a single-occupied sites~\cite{TKLee03}.
This doublon effect leads to a spin singlet states between the two planes.

In an undoped bilayer system, both planes have same amount of carriers due to the self-doping, although one type of carrier is hole and the other is electron.
Our results show that in both electron- and hole-doped planes, AFM and SC coexist.
Other authors studied the undoped case with no interlayer hopping~\cite{Han06,Ribe06}.
We examine the doped case with holes as well.
In the doped case, the numbers of carriers in each plane becomes imbalanced,~i.e.~doublon density decreases and holon density increases with hole doping.
As a result, the magnitudes of order parameters become more dissimilar compared to the undoped case.
Two FS's in the self-doped bilayer look like a composite of hole- and electron-doped cuprates.
However, we cannot find the FS splitting in the nodal direction, since one band becomes very flat due to the AFM orders.

This paper is organized as follows. In Sec.~\ref{sec2}, we introduce the bilayer $t$-$t^\prime$-$J$ model with an interlayer hopping and a site potential, and present the slave-boson mean-field scheme.
In Sec.~\ref{sec3}, we discuss self-consistent mean-field solutions for both undoped and doped cases of self-doped bilayer system.
Coexistent phase of SC and AFM orders are discussed from the viewpoint of doping and charge imbalance.
FS and dispersion relation of spinons are shown in the AFM ordered phase.
In Sec.~\ref{sec4}, we will give summary and discussion.

\section{Model and method} \label{sec2}

The model we apply to study the self-doped bilayer system is
written as:
\begin{eqnarray}
  H & = & H_\parallel + H_W + H_\perp ,\label{Hamil}\\
  H_\parallel & = & \sum_{l=1,2}
    \biggl[\Bigl( -t\sum_{\langle ij\rangle,\sigma}
        c_{i\sigma}^{(l)\dagger} c_{j\sigma}^{(l)}
    -t^\prime \sum_{(ij),\sigma}
        c_{i\sigma}^{(l)\dagger}c_{j\sigma}^{(l)}
            +h.c. \Bigr)\nonumber\\
    &&+ J\sum_{\langle ij\rangle}
        \Bigl(S_i^{(l)}\cdot S_j^{(l)} -
            \frac{1}{4}n_i^{(l)}n_j^{(l)}\Bigr)
    - \mu \sum_{i,\sigma} n_{i\sigma}^{(l)}\biggr],\\
  H_W & = & W\sum_i \left(n_{i}^{(1)} - n_{i}^{(2)}\right) ,\\
  H_{\perp} & = &  \sum_{i,j,\sigma}\left( -t_{\perp ij}
    c_{i\sigma}^{(1)\dagger} c_{j\sigma}^{(2)}+ h.c.\right), \label{inter}
\end{eqnarray}
where $c_{i\sigma}^{(l)}$ ($c_{i\sigma}^{(l)\dagger}$) is the
electron annihilation (creation) operator with spin $\sigma$ at
site $i$ in the $l$-th plane.
The electron number in each plane is denoted by $n_{i}^{(l)}=\sum_\sigma
c_{i\sigma}^{(l)\dagger} c_{i\sigma}^{(l)}$, and the averaged electron
density is defined as, $n\equiv(n^{(1)} + n^{(2)})/2$.
The signs, $\langle ij\rangle$ and $(ij)$, run over nearest- and next-nearest
neighbor sites, respectively.
The chemical potential $\mu$ and the site potential $W$ control the charge imbalance.
Below, we take $J/t=1/3$ and $t^\prime/t=-0.4$.

The interlayer hopping in Eq.~(\ref{inter}) has the dispersion
relation in the momentum space,
$
\varepsilon_{\perp,k}
    =(t_{\perp}/4)\left(\cos k_{x}-\cos k_{y}\right)^{2},
$
where $t_\perp$ is the amplitude without renormalization~\cite{Chak93,Ande95,Liech96,Feng01,Chua01}.

We treat Hamiltonian $(\ref{Hamil})$ in the slave-boson mean-field theory.
The electron operator is represented as,
$c_{i\sigma}^{(l)} = f_{i\sigma}^{(l)} h_i^{(l)\dagger} + \sigma f_{i\bar\sigma}^{(l)\dagger} d_i^{(l)}$,
 with $h_i^{(l)}$ and $d_i^{(l)}$ being the bosonic holon and doublon operators, respectively~\cite{Barn76,Isaw87}.
The fermionic spinon operator is denoted by $f_{i\sigma}^{(l)}$.
In the self-doped case, we assume that one plane is hole-doped and
the other is electron-doped.
For the hole-doped plane, as there is no
doublon, the electron operator can be expressed as,
$c_{i\sigma}^{(l)} = f_{i\sigma}^{(l)} h_i^{(l)\dagger}$,
with the constraint,
$h_i^{(l)\dagger} h_i^{(l)} + \sum_\sigma f_{i\sigma}^{(l)\dagger} f_{i\sigma}^{(l)} =1$,
while for the electron-doped plane, as there is no holon,
the electron operator can be expressed as,
$c_{i\sigma}^{(l)} = \sigma f_{i\bar\sigma}^{(l)\dagger} d_i^{(l)}$,
with the constraint,
$d_i^{(l)\dagger} d_i^{(l)} + \sum_\sigma f_{i\sigma}^{(l)\dagger} f_{i\sigma}^{(l)} =1$.
Since we are interested in the electronic states at low
temperatures, the boson condensation is assumed in each plane,
i.e.,
$\langle h_i^{(l)}\rangle = \langle h_i^{(l)\dagger}\rangle = \sqrt{\delta_h^{(l)}}$, and
$\langle d_i^{(l)}\rangle = \langle d_i^{(l)\dagger}\rangle = \sqrt{\delta_d^{(l)}}$,
where $\delta^{(l)}_h$ and $\delta^{(l)}_d$ are the holon and doublon
densities.

To decouple the Hamiltonian, we introduce the order parameters in
the electron- and the hole-doped planes as,
$   \Delta_\eta^{(l)} = \langle f_{i\downarrow}^{(l)}
    f_{i+\eta\uparrow}^{(l)} -
    f_{i\uparrow}^{(l)} f_{i+\eta\downarrow}^{(l)}\rangle
$,
$   \chi_\eta^{(l)} =
    \langle f_{i\sigma}^{(l)\dagger}f_{i+\eta\sigma}^{(l)}\rangle
$,
$
  m^{(l)} = (-1)^{(l+1)} \langle n_{i\uparrow}^{(l)} - n_{i\downarrow}^{(l)}\rangle
$,
where $\eta=x,y$ indicates the nearest-neighbor sites.
Although the magnetic order in real materials may be quite complicated, we only
consider the commensurate antiferromagnetic orders for simplicity.
As the interlayer hopping may induce a weak AFM correlation
between the two planes, the staggered AFM order has a sign
difference between the two planes in our definition. All
parameters are assumed to be real and the SC pairing symmetry is
$d$-wave.

The Hamiltonian $(1)$ based on the above treatment is decoupled
in the momentum space as follows:
\begin{eqnarray}
  H_{\rm MF} &=& \sum_{l,k,\sigma }
        \bigl(
            \varepsilon _{k}^{(l)}f_{k\sigma}^{(l)\dagger }f_{k\sigma }^{(l)}
            +\varepsilon_{k+Q}^{(l)}f_{k+Q\sigma}^{(l)\dagger }f_{k+Q\sigma }^{(l)}
        \bigr)\nonumber\\
    &-&\frac{1}{2}J\sum_{l,k} \Delta ^{(l)}\eta_{k}
        \bigl(
            f_{-k\downarrow }^{(l)}f_{k\uparrow }^{(l)}
            -f_{-k+Q\downarrow}^{(l)}f_{k+Q\uparrow}^{(l)}+h.c.
        \bigr) \nonumber\\
    &-&J\sum_{l,k,\sigma } (-1)^{(l+1)}m^{(l)}\sigma
        \bigl(
            f_{k\sigma}^{(l)\dagger }f_{k+Q\sigma }^{(l)}   +h.c.
        \bigr) \nonumber\\
    &-&\sum_{k,\sigma }\sqrt{\delta_h^{(1)}\delta_{d}^{(2)}}
        \varepsilon _{\perp ,k}\cdot\sigma \nonumber\\
    &&\times
        \bigl(
            f_{-k\overline{\sigma }}^{(2)}f_{k\sigma }^{(1)}
            +f_{-k+Q\overline{\sigma}}^{(2)}f_{k+Q\sigma }^{(1)} +h.c.
        \bigr),\nonumber\\
    &+& JN\sum_{l}
        \bigl(
            \Delta ^{(l)2}+\frac{1}{2}\chi ^{(l)2}+\frac{1}{2}
            m^{(l)2}+\frac{1}{2}n^{(l)2}
        \bigr), \label{MF}
\end{eqnarray}
where $\gamma_k=2(\cos k_x +\cos k_y)$, $\eta_k = 2(\cos k_x -\cos
k_y)$, $\zeta_k=4\cos k_x  \cos k_y$, and $k$ runs over the
magnetic Brillouin zone with $|k_x\pm k_y|\leq\pi$.
$Q=(\pi,\pi)$ is the magnetic vector and $N$ is the total number
of lattice sites. $\sqrt{\delta^{(1)}_{h}\delta^{(2)}_{d}}$ is the
renormalization factor of $t_\perp$\cite{Mori06}.
Here, we assumed the $l$=1 (2) is the hole (electron) doped plane.

We note that as seen in the fourth term in Eq.~(\ref{MF}), the interlayer
hopping in self-doped bilayer system may induce an interlayer
singlet-paring, which can be defined as,
$
\Delta_p = \langle f_{i\downarrow}^{(1)} f_{i\uparrow}^{(2)} -
f_{i\uparrow}^{(1)} f_{i\downarrow}^{(2)}\rangle.
$

The momentum dependence of dispersions is given by
$
\varepsilon _{k}^{(1)}
    =-(t\delta^{(1)}_{h}+\frac{1}{4}J\chi^{(1)})\gamma_{k}
        -t^{\prime }\delta^{(1)}_{h}\zeta _{k}-(\mu + Jn_{s}^{(1)}-W)
$,
$
\varepsilon _{k}^{(2)}
    =   (t\delta^{(2)}_{d}-\frac{1}{4}J\chi^{(2)})\gamma _{k}
        +t^{\prime }\delta^{(2)}_{d}\zeta _{k}-(\mu + Jn_{s}^{(2)} + W)
$,
where
$
n_{s}^{(1)}=\sum_{\sigma}f^\dag_\sigma f_\sigma
$
and
$
n_{s}^{(2)}=\sum_{\sigma}g^\dag_\sigma g_\sigma
$
are spinon densities in plane 1 and 2, respectively.
For the given total electron number $n$ and site potential $W$,
the mean-field parameters $\Delta^{(l)}$, $\chi^{(l)}$ and
$m^{(l)}$, the charge density in each plane $\delta^{(1)}_h$
($\delta^{(2)}_d$) and the chemical potential $\mu$ are
self-consistently determined in numerical calculations.

\section{Results and discussion} \label{sec3}

\subsection{Self-doped bilayer at half-filling ($n$=1)} \label{half}

\begin{figure}[t]
\begin{center}
\includegraphics[width=9.5cm,height=10cm]{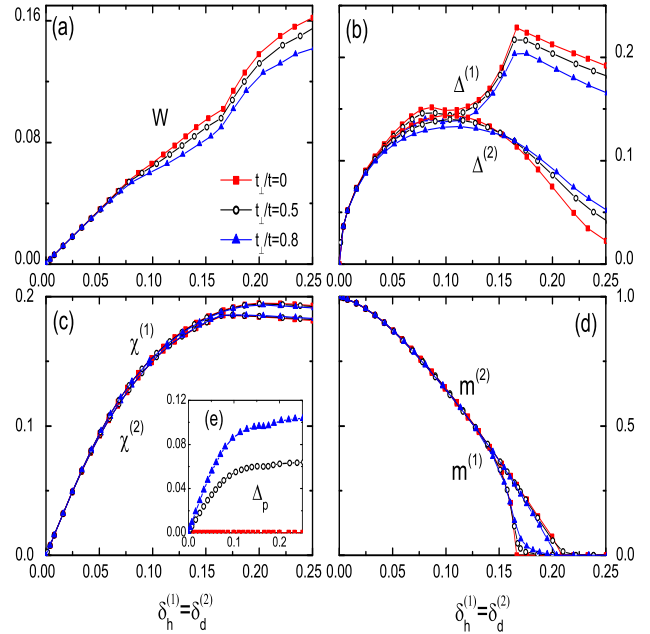}
    \caption{
The order parameters in the undoped case with
$t'/t=-0.4$. The antiferromagnetic (AFM) and the superconducting (SC) orders coexist in both planes.
The following properties in electron- and hole-doped planes are plotted as functions of holon (doublon) density;
(a) site potential, $W$,
(b) $d$-wave pairing amplitude, $\Delta^{(l)}$,
(c) the uniform bond order, $\chi^{(l)}$,
(d) AFM order parameter, $m^{(l)}$, and
(e) interlayer singlet-paring, $\Delta_p$.
Three lines in each figure correspond to different value of interlayer hopping given by,
$t_\perp/t$=0.0 (square),
$t_\perp/t$=0.5 (circle),
$t_\perp/t$=0.0 (triangle).
    } \label{fig1}
\end{center}
\end{figure}

First, we focus our study on the undoped case, i.e., $n=1$.
In this case, the holon density in hole-doped plane is equal to the doublon density in electron-doped plane, i.e.,
$\delta_h^{(1)}=\delta_d^{(2)}$.
Figure~\ref{fig1} shows the results of the $d$-wave pairing amplitude ($\Delta^{(l)}$),
the uniform bond order parameter ($\chi^{(l)}$), the AFM order parameter ($m^{(l)}$), the site potential ($W$) and the interlayer singlet pairing amplitude ($\Delta_p$) as functions of the holon (doublon)
density $\delta_h^{(1)}$ ($\delta_d^{(2)}$) for various values of the interlayer hopping parameter ($t_\perp$).
$\Delta^{(l)}$, $\chi^{(l)}$, and $m^{(l)}$ depend very weakly on $t_\perp$,
particularly for small $W$ (small $\delta_h^{(1)}$ and $\delta_d^{(2)}$).
$\Delta_p$ increases with $t_\perp$. When $t_\perp=0$,
$\Delta_p=0$. Fig.~\ref{fig1} (a) shows the relation between $W$ and
$\delta_h^{(1)}$ ($\delta_d^{(2)}$). When $W=0$, there is no charge imbalance
between the two planes, i.e., $\delta_h^{(1)}=\delta_d^{(2)}=0$. In this case,
both planes are at half-filling, with $m^{(l)}=1$ and
$\Delta^{(l)}=\chi^{(l)}=0$. The ground state is an AFM insulator.
When $\delta_h^{(1)}=\delta_d^{(2)}=0$, $\Delta_p$ becomes zero, that is, the
planes are decoupled regardless of the interlayer hopping and the
planes are coupled only for finite $\delta_h^{(1)}$
($\delta_d^{(2)}$)~\cite{Mori06}. When $W$ increases, the charge densities
$\delta_h^{(1)}$ and $\delta_d^{(2)}$ increase from zero. The staggered AFM
magnetization decreases with $\delta_h^{(1)}$ ($\delta_d^{(2)}$), while the
$d$-wave paring amplitude ($\Delta^{(l)}$) and the uniform bond
order parameter ($\chi^{(l)}$) both increase. $m^{(1)}$ and
$m^{(2)}$ are almost the same in the region $0<\delta_h^{(1)}$
,$\delta_d^{(2)}\lesssim 0.15$, and then $m^{(1)}$ decreases faster than
$m^{(2)}$ and vanish at around $\delta_h^{(1)}=\delta_d^{(2)}\sim 0.2$.
It is seen that for $0<\delta_h^{(1)}$ ,$\delta_d^{(2)}\lesssim 0.2$, both
electron- and hole-doped planes are the coexistent state of AFM
and SC. When AFM order vanishes, both planes are superconducting.

Here, we mention yet another solution,
where $\delta_h^{(1)}=\delta_d^{(2)} \neq 0$ even for $W=0$, and $m^{(2)} \gg m^{(1)} \sim 0$.
This may be a possible {\it phase separation}, where electrons are spontaneously transferred from the hole-doped plane to the electron-doped one to gain an energy  of magnetic exchange interaction without the site potential.
Since this solution is found in a limited (unphysical) parameter region,
we do not discuss below.

\subsection{Hole-doped case ($n<$1)} \label{doped}
\begin{figure}[t]
\includegraphics[width=9.5cm,height=10cm]{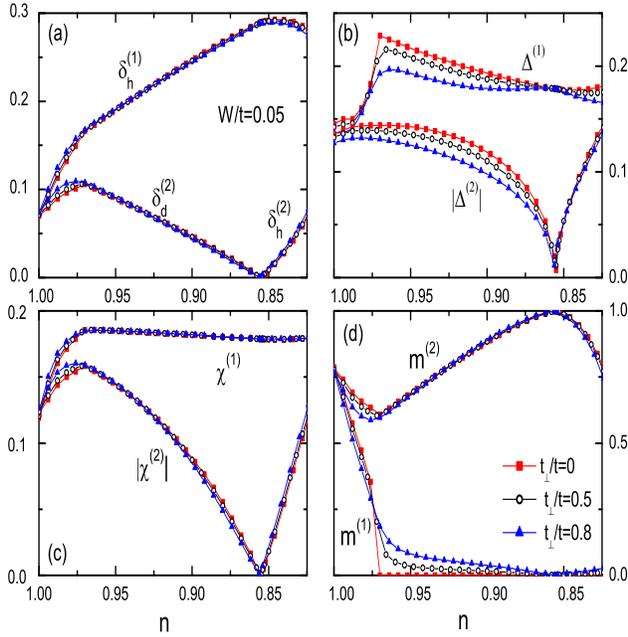}
    \caption{
The order parameters in the doped case with $t'/t=-0.4$
and $W/t=0.05$.
The following properties in electron- and hole-doped planes are plotted as functions of total electron
number $n$;
(a) holon (doublon) density,
(b) $d$-wave pairing amplitude, $\Delta^{(l)}$,
(c) the uniform bond order, $\chi^{(l)}$, and
(d) AFM order parameter, $m^{(l)}$.
Three lines in each figure correspond to different value of interlayer hopping given by,
$t_\perp/t$=0.0 (square),
$t_\perp/t$=0.5 (circle),
$t_\perp/t$=0.0 (triangle).
    } \label{fig2}
\end{figure}

Next we investigate the doped case in the self-doped bilayer system.
In the doped case, i.e., $n\not = 1$, the holon density
$\delta_h^{(1)}$ is not necessarily equal to doublon density
$\delta_d^{(2)}$. Figure~\ref{fig2} shows the results of
$\delta_h^{(1)}$ ,$\delta_d^{(2)}$, $\Delta^{(l)}$, $\chi^{(l)}$,
and $m^{(l)}$ as functions of the total electron density $n$ for a
given $W/t=0.05$. From Fig.~\ref{fig2} (a), we see that for small
doping of holes, $0.97<n<1$, both $\delta_h^{(1)}$ and
$\delta_d^{(2)}$ increase. This means that holes first go into
hole-doped plane, while some electrons are transferred from
hole-doped plane to electron-doped plane. The kinetic energy gains
in this case. Upon further doping of holes into the system, holes go
into both planes, and $\delta_h^{(1)}$ increases while
$\delta_d^{(2)}$ decreases. Due to the change of charge density, the
staggered AFM magnetization in hole-doped plane ($m^{(1)}$)
decreases while $m^{(2)}$ increases. For $0.86\lesssim n<1$, both
electron- and hole-doped planes are the coexistent state of AFM and
SC. For $0.86\lesssim n \lesssim 0.97$, the AFM order in hole-doped
plane is small, and $m^{(1)}$ decreases fast with $t_\perp$. When
$t_\perp=0$, $m^{(1)}$ becomes zero in this region. At a critical
point $n\sim 0.86$, doublon vanishes ($\delta_d^{(2)}=0$), and
electron-doped plane goes into AFM insulator phase; simultaneously
$m^{(1)}$ becomes zero, and hole-doped plane goes into the
superconducting phase. Above the critical point ($n<0.86$), both
planes are hole-doped.

\begin{figure}[t]
\includegraphics[width=8.5cm,height=8.5cm]{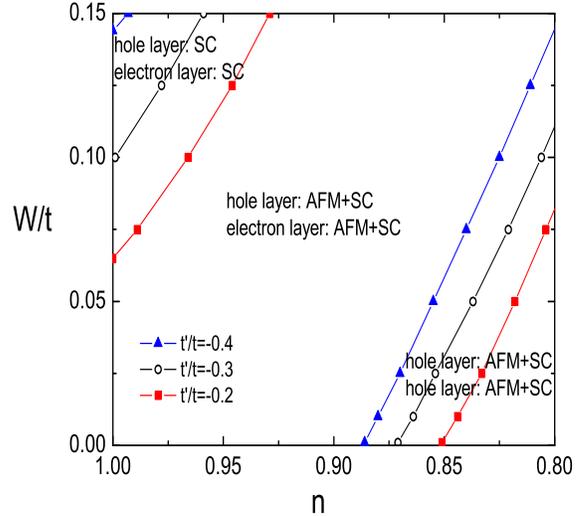}
    \caption{
The phase diagram in the $W$-$n$ plane for $t_\perp/t=0.5$.
Each abbreviation means the followings:
AFM, the antiferromagnetic phase:
SC, $d$-wave superconducting phase:
AFM+SC, the coexistent phase of AFM and SC.
For large $n$, both planes are hole doped. Different lines correspond to different values of $t'/t$ as,
-0.2, -0,3 and -0.4 from bottom to top.
    } \label{fig3}
\end{figure}

So far we have presented results for both undoped and doped cases.
Now we discuss the phase diagram. Fig.~\ref{fig3} shows the phase diagram
in the $W$-$n$ plane for $t_\perp/t=0.5$. The phase diagram is
divided into three parts. For small doping ($n$ close to $1$) and
large site potential $W$, the charge imbalance is large and both
planes are SC; for small $W$ and small $n$, both planes are
hole-doped; for intermediate $W$ and $n$, both electron- and
hole-doped planes are the coexistent state of AFM and SC. In the
undoped case ($n=1$), with increasing $W$, $\delta_h^{(1)}=\delta_d^{(2)}$
increases and there is a transition from coexistent state of AFM
and SC to SC in both planes.
In the doped
case with $n=0.85$, both planes are hole-doped for small $W$. When
$W$ increases above a critical value of $W_C$, electrons move from
hole-doped plane to electron-doped plane and both planes are the
AFM and SC coexistent state due to self-doping.

\subsection{Spinon Fermi surfaces and dispersion relation}
\begin{figure}[t]
\includegraphics[width=9.5cm]{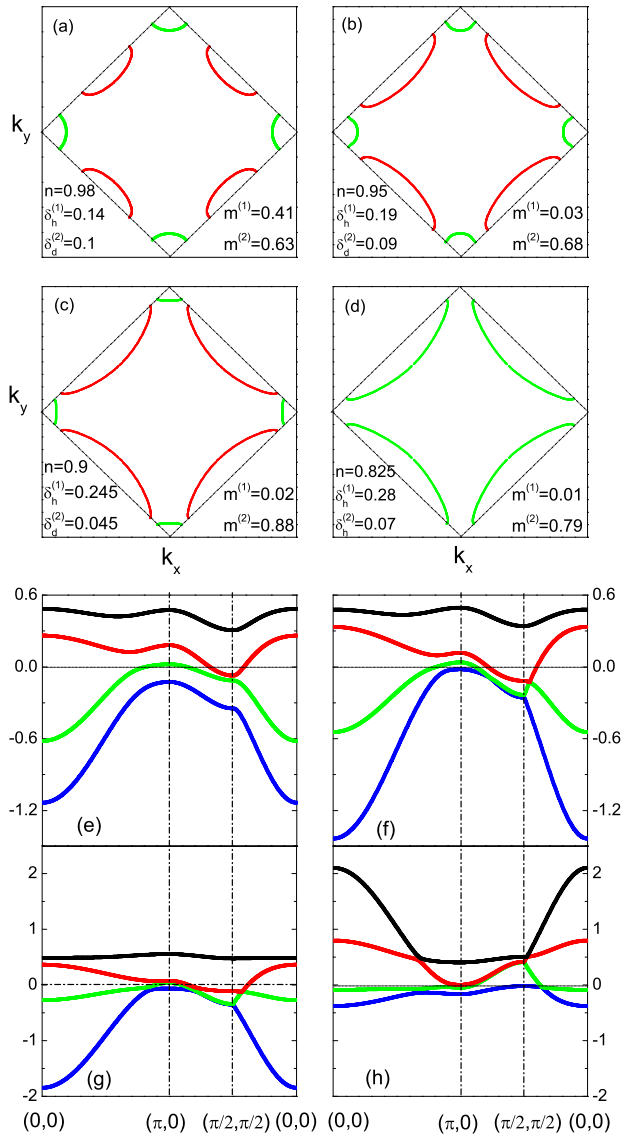}
\caption{
FS's and dispersion relations of self-doped bilayer system;
(a) and (e) $n$=0.98, $\delta_{\rm h}^{(1)}$=0.130, $\delta_{\rm d}^{(2)}$=0.090;
(b) and (f) $n$=0.95, $\delta_{\rm h}^{(1)}$=0.180, $\delta_{\rm d}^{(2)}$=0.080;
(c) and (g) $n$=0.90, $\delta_{\rm h}^{(1)}$=0.225, $\delta_{\rm d}^{(2)}$=0.025;
(d) and (h) $n$=0.85, $\delta_{\rm h}^{(1)}$=0.25, $\delta_{\rm h}^{(2)}$=0.05.
In these figures, $W/t$=0.05 and $t_\perp/t$=0.5 are fixed, and SC order parameters are imposed to be zero.
In~(d) and (h), both two planes are in hole-doped regions.
} \label{fig4}
\end{figure}

As for the {\it asymmetry} between the hole- and the electron-doped
cuprates, one of distinguished observations is the FS pocket, which
is located around the nodal region in hole-doped
cuprates\cite{Ino02,Yosh03,Ronn03,Shen05} and the anti-nodal region
in electron-doped ones~\cite{Armi02}. It is found that this {\it
asymmetry} originates from the different signs of $t'$ and
$t''$~\cite{TKLee97,TKLee03,Tohy00,Tohy04}. On the other hand,  the
multilayered cuprates doped with holes show the interlayer
splittings of
FS~\cite{Chak93,Ande95,Liech96,Feng01,Chua01,Mori06,Mori02,Koik06}.
In the nodal direction, the splitting means the charge imbalance
between IP and OP, while those around the anti-nodal regions are
ascribed to a magnitude of interlayer hopping renormalized by the
charge imbalance~\cite{Mori06}. Interesting is that the two
asymmetric planes are combined by the interlayer hopping in the
self-doped bilayer system.

In Fig.~\ref{fig4}, the spinon FS's and dispersion relations of self-doped bilayer systems are plotted for some doping rates in the AFM phase.
Details of parameters are included in the caption of Fig.~\ref{fig4}.
Near the half-filling given by, $n$=0.98, $\delta_{\rm h}^{(1)}$=0.130, and $\delta_{\rm d}^{(2)}$=0.090, two FS pockets appear in the nodal and the antinodal regions shown in Fig.~\ref{fig4} (a).
It looks like a composite of hole- and electron-doped cuprates.
As shown in Fig.~\ref{fig4} (e), four dispersion relations are separated each another due to the large AFM moments, $m^{(1)}$=0.47 and $m^{(2)}$=0.68.
With increasing hole density in the bilayer as $n$=0.95 and 0.9, the hole-doped like FS becomes larger as shown in Fig.~\ref{fig4} (b) and (c), and the AFM moment in the hole-doped plane markedly becomes small, $m^{(1)}$=0.05 and 0.01.
As a result, the separation of four bands close on each another as shown in Figs.~\ref{fig4} (f) and (g).
On the other hand, since the AFM moment in the electron-doped plane still large, $m^{(2)}$=0.73 and 0.95, two among four bands become quite flat.
Finally, for $n$=0.85, both plane becomes hole-doped.
Interesting is that we cannot find the interlayer splitting in the nodal direction as shown in Fig.~\ref{fig4} (d), although it is found in the normal metallic phases~\cite{Mori06,Mori02}.
The missing of FS splitting is caused by the AFM moment in the 2nd plane, which make a band flat as shown in Fig.~\ref{fig4} (h).

\section{Summary and discussion} \label{sec4}

We have studied the bilayer self-doped cuprates  by using the
slave-boson mean-field theory. Each plane is described by the
$t$-$t^\prime$-$J$ model, and the interlayer hopping and a site
potential are included. In an undoped bilayer system, both planes
have same amount of carriers due to the self-doping, although one
type of carrier is hole and the other is electron. Our results show
that in both electron- and hole-doped planes, AFM and SC coexist. In
the doped cases with holes, the numbers of carriers in each plane
becomes imbalanced,~i.e.~doublon density decreases and holon density
increases with hole doping. The magnitudes of order parameters
become more dissimilar compared to the undoped case. At some
critical doping of holes, the doublon disappears and the electron
layer becomes an insulator. This effect might be useful for a p-n
junction \cite{Hanamura} made of the electron- and hole- doped
layers if we could control the doping near this value. Two FS's in
the self-doped bilayer look like a composite of hole- and
electron-doped cuprates. However, we cannot find the FS splitting in
the nodal direction even in the bilayer system, since one band
becomes very flat due to the AFM orders.

In the ARPES experiment on F0234, two FS's surrounding the
($\pi$,$\pi$) point and the FS's splitting in the nodal direction
are observed. The doping rate in each plane may be optimal or
overdoped. In such a case, the FS should enclose the ($\pi$,$\pi$)
point like as Bi-compounds, not like as a pocket. On the other hand,
the NMR study observed the magnetic moment, which could not exist
around the optimum doping region. Although our model is a bilayered
system, it involves essential points of multilayered cuprates. In
addition to the bilayered system, the four-layered $t$-$t'$-$J$
model was examined to find the FS splitting in the nodal region.
However, we could not find it in the self-doped four-layered system,
while it was found in the four hole-doped layers. The contradiction
between experiment and theory remains to be resolved in the future.

\begin{acknowledgments}
The authors would like to thank G. Khaliullin for valuable discussions.
This work was supported by a Grand-in-Aid for Scientific Research on
Priority Areas and the NAREGI Nanoscience Project from MEXT, and CREST.
The authors thank the Supercomputer Center, ISSP, University of Tokyo.
\end{acknowledgments}


\end{document}